\documentclass[twocolumn,showpacs,preprintnumbers,amsmath,amssymb,floatfix]{revtex4}
\usepackage{graphicx}

\usepackage[center]{subfigure}
\usepackage{bigints}
\usepackage{cancel}
\usepackage{amsmath}
\usepackage{mathbbol}
\usepackage[center]{subfigure}
\usepackage{tikz}
\usepackage{standalone}
\usepackage{stackengine}
\usepackage{xr}
\usepackage{cleveref}

\usepackage{mathtools}

\begin{document}
\author{Tarek Tohme$^{1}$, Martin Grant$^{2}$, and Sara Najem$^{3}$}

%
\affiliation{$^1$ Electrical and Computer Engineering Department, American University of Beirut, Beirut 1107 2020, Lebanon}

\affiliation{$^2$ Physics Department, Rutherford Building, 3600 rue University, McGill University,
Montreal, Qu\'ebec, Canada H3A 2T8}

\affiliation{$^3$ Physics Department, American University of Beirut, Beirut 1107 2020, Lebanon}


\date{\today}

\begin{abstract}
We look at buildings' competition over space in cities following 
 {the distribution $p(r)$} of the perimeters $r$ of the buildings' circumscribing ellipses.  $p(r)$  {is shown to follow} a power-law behavior beyond a critical threshold of the density of the built environment. 
In this regime, $p(d)$, where $d$ is the distance to the nearest competitor, defined to be a building with a larger $r$, bifurcates with the buildings' number $n$. This reveals two different  {competition} laws: one which is linked  {to} spatial homogeneity and segregation, as opposed to another favoring  spatial diversity and intermixing  {between buildings with different sizes}. 

 \end{abstract}
\title{Buildings as Species: Competition and Scaling Rules in Cities}

\maketitle
\section{Introduction}

 {The} interplay between size and abundance 
in urban environments is manifested through the emergence of scaling laws or self-similar patterns relating growth metrics to size \cite{volkov2003neutral,Simini:2010jza,Batty:2008bsb,wiens1989spatial,depersin2018global,barthelemy2016structure,barenblatt2003scaling,west1997general,west1999fourth,barthelemy2008modeling,louf2013modeling,riascos2017universal,goh2016complexity}. However, this systems-of-cities approach blurs the spatial structure within cities and zooms out to study them as points in the size versus population/income/employment space \cite{bettencourt2007growth}. 

Conversely to this inter-cities approach there are numerous measures to assess the city's internal spatial organization, and the literature abounds with examples examining the elementary processes governing these spatial mechanisms \cite{strano2012elementary,makse1995modelling,lammer2006scaling}.  These measures serve as proxies to determine its level of connectivity, resilience, accessibility, sustainability, and livability \cite{boeing2018measuring,molinero2019geometry,anas1998urban,wilson2014complex,keuschnigg2019urban,brelsford2018toward}. These span: $(i)$ information theoretic metrics, such as Shannon's information entropy, which reveal the spatial embedding of urban design, $(ii)$ measures of self-similarity and fractal dimensions, which uncover the interdependence between physical structure and topological arrangement, $(iii)$ geometrical and topological characteristics of urban networks. Notions like atomic-scale structure, borrowed from condensed matter physics, are also used as identifiers to provide additional insight into a city's geometrical patterns and texture \cite{sobstyl2018role,najem2019streets,najem2017solar}. 

The ensuing spatial order of buildings, which is of particular interest to us in this work, and is quantified by the above measures, was shown to be associated with the interaction of microeconomic forces, urban design, as well as geometrical constraints \cite{anas1998urban,schlapfer2015urban}.  City growth, an example of what Mitzenmacher terms `multiplicative processes', can generate lognormal or power-law distributions and scaling laws  \cite{mitzenmacher2004brief,clauset2009power}. When resources are limited, sustainable distributions of city components,  buildings in our case, seem to be those which permit the presence of a small number of large buildings as opposed to a large number of small ones.
Particularly, the buildings' size distribution, which is a power-law, is a signature of such an evolving competitive process \cite{Batty:2015ida}. 

In this context, our study of intra-city dynamics, inspired by \cite{Simini:2010jza} and by similarities of form between urban systems and  ecosystems drawn by Wilson \cite{wilson2006ecological,wilson2007boltzmann}, investigates competition rules between buildings' species. The analogy we draw here between forests and buildings is substantiated by the increased demand for space concomitant with urban sprawl, population growth, densification, accessibility to resources, and their intensification. 
Speciation, here, is defined as footprint perimeter, and a competitor is understood as a building with a larger perimeter. Competition is studied through the distribution of the distance to the nearest competitors. A city where competition for space is weak entails a mixture of building species across the urban fabric, whereas fierce competition tends to homogenize species distribution into segregated neighborhoods. We also compute buildings' orientations and consequently the entropy associated with their directionality,  and the length of the roads. Relying on these metrics we identify two distinct scaling regimes characterizing two different ``predatory rules" between buildings.

\section{Methods}
\begin{figure}[!htp]%
 \center
 \vspace{-1.8cm}
    \includegraphics[width=0.3\textwidth]{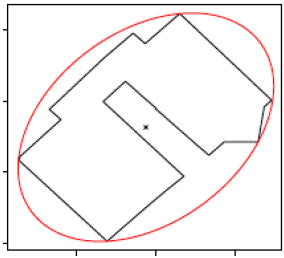} 
    \vspace{-1.8cm}
  \caption{ In red is ellipse circumscribing the building's footprint shown in black.}
  {\label{fig:1}}
\end{figure}

The distributions $p(r)$, where $r$ is the perimeter of the circumscribing ellipse of a building's footprint, shown in Figure \ref{fig:1}, were followed for a sample of 1,500 cities in the US using OpenStreetMap data, which is to the best of our computational capabilities. 
 More specifically, species are distinguished by the perimeter of the ellipse around the building's footprint $r$  \cite{Simini:2010jza}. It is a preferable alternative to the footprint perimeter since the latter does not necessarily define a no contact space as shown in Figure \ref{fig:1}.
 
The size distribution of buildings is given by:

  \begin{equation}\label{sizedist}
p(r) \propto (r/r_{min})^{-\alpha},
\end{equation}
 where $r_{min}$ and $  \alpha$ are respectively the lower cutoff and the exponent of the cumulative distribution function $P(r)$ characterizing each city. They were computed using the poweRlaw package in R.  In what follows, we restrict our analysis to the cities whose $P(r)$ and thus $p(r)$  {of Eq. \ref{sizedist}} pass the power law test; they correspond to cities with $n$ buildings' greater than $n_c$, where $n_c = e^{4.7}$ is the critical number of buildings below which the distribution fail the power-law test. 

\begin{figure}[!htp]%
 \center
    \includegraphics[width=0.5\textwidth]{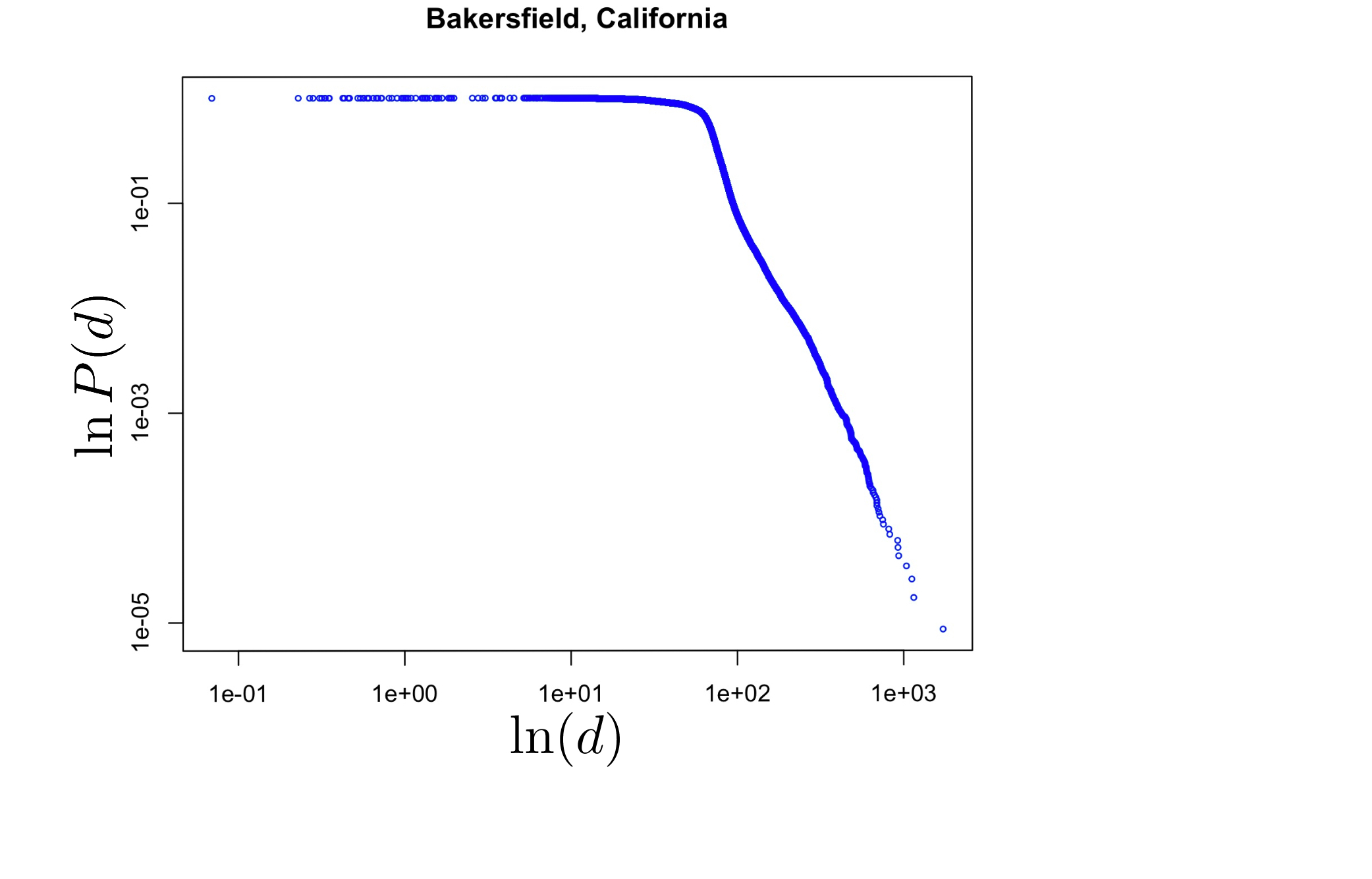} 
  \caption{ The cumulative distribution $P(d)$ of Bakersfield, California on a log-log scale.}
  {\label{examplePofD}}
\end{figure}

 {Given the inter-competitors distances $d$, taken to be here the shortest distance to a building with a radius larger than $r$}, allows us to calculate $p(d|r)$ the conditional distribution and subsequently the non-conditional distributions of competitors $p(d)$.  {Beyond $n_c$, $p(d)$ followed a power law-behavior given by:}
  \begin{equation}\label{distributionpd}
p(d) \propto (d/d_{min})^{-\gamma},
\end{equation}
where $d_{min}$ is the lower-cutoff distance of the power law. These exponents were retrieved by applying the power law test on the cumulative distribution $P(d)$,  {a sample of which is shown in Fig. \ref{examplePofD} on a logarithmic scale.}

We note  that the combination  $\phi = d_{min}^{\gamma}$ completely characterizes $p(d)$, which we follow as function of $n$ and $L$, where the latter is the street length. Additionally, the effect of city size is measured by renormalizing $d_{min}$ by $L$, since the street length is a proxy to inter-buildings' distance.
For each city we additionally compute its total streets length $L$, its buildings' size and orientation entropies respectively given by: 

\begin{equation}\label{entropy}
{S} = -\sum_i^N p_i \log{p_i}, 
\end{equation}
when $p_i$ is the probability of a buildings to have a size $i$, the definitions corresponds to size entropy denoted by $\mathcal{S}_{size}$ in what follows, while when $p_i$ is the probability of a building to be oriented along direction $i$ it corresponds to the entropy of orientations denoted by $\mathcal{S}$.

Further, we looked at the average size entropy $\bar{\mathcal{S}}_{size}$ as a function of $n$ and $L$.  {The rationale behind computing the entropy is that its change with respect to $n$ is known to be related to the chemical potential $\mu$, which measures the cost of constructing an additional building, and thus any abrupt change in entropy is a signature of a phase transition, which we will explore in the results section. }

\section{Results and Discussion}
\begin{figure}[!htp]%
 \center
    \includegraphics[width=0.6\textwidth]{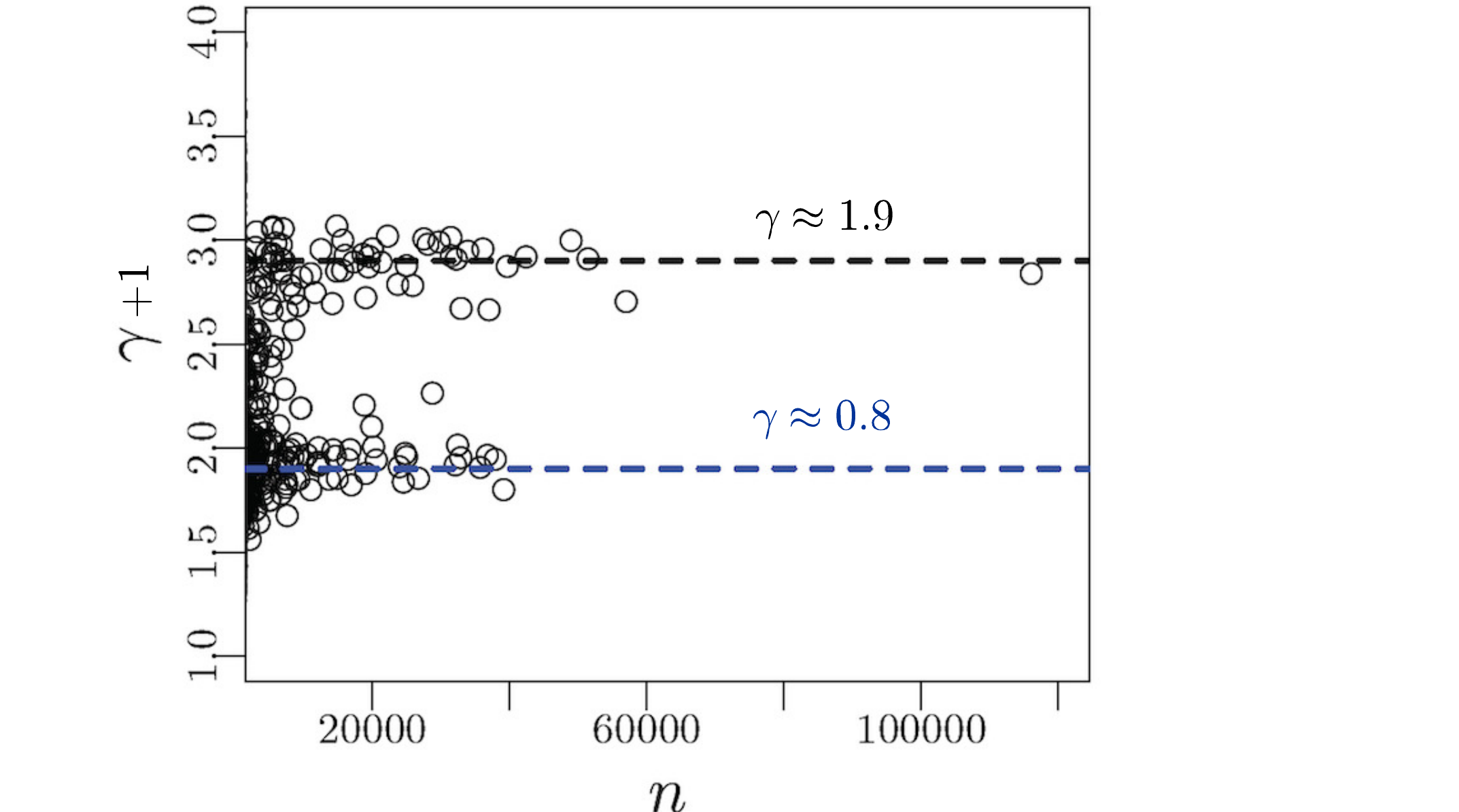} 

  \caption{ The values of $\gamma + 1$ plotted against the number of buildings.}
  {\label{gammavsnbbui}}
\end{figure}

\begin{figure}[!htp]%
  \begin{subfigure} {\label{fig:dmingamvsn}}
    \centering\includegraphics[width=10cm]{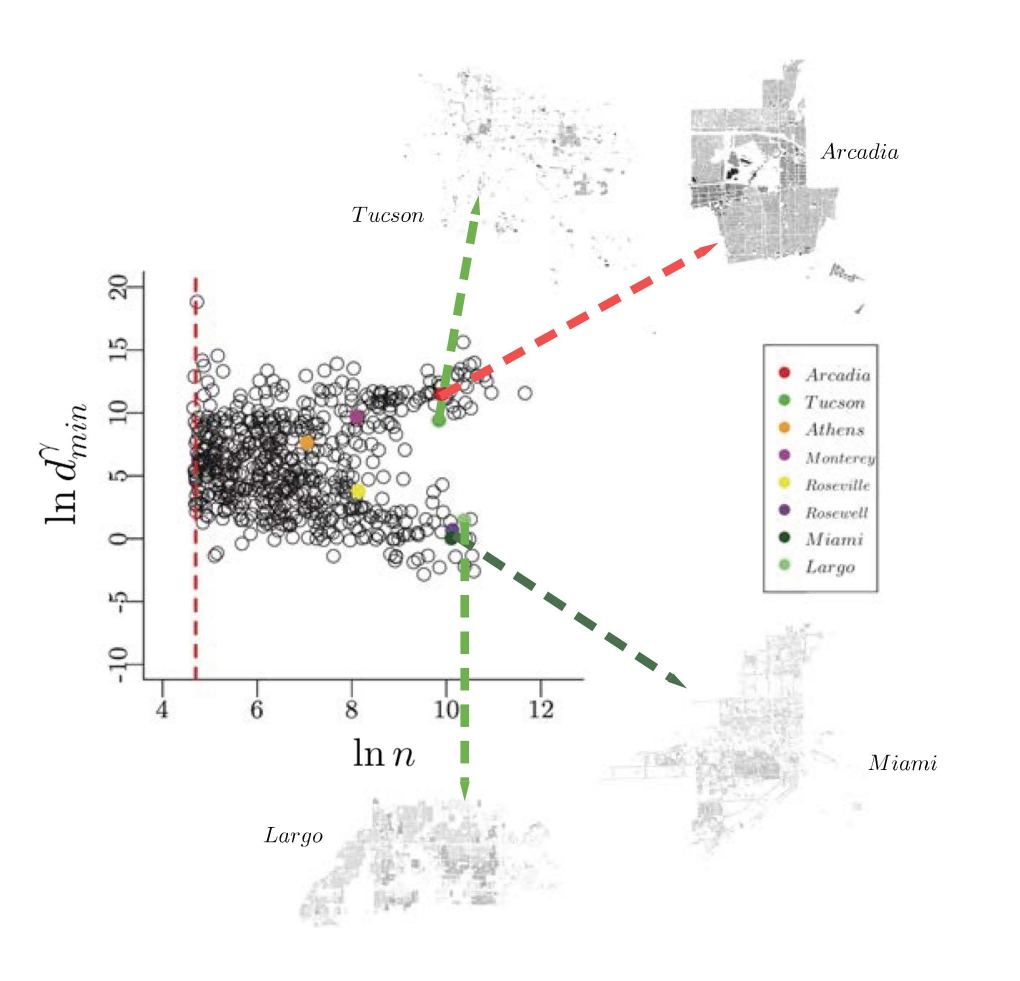}

  \end{subfigure}
    \begin{subfigure}{\label{fig:DistVsNbLnormalized}}
    \centering\includegraphics[width=7cm]{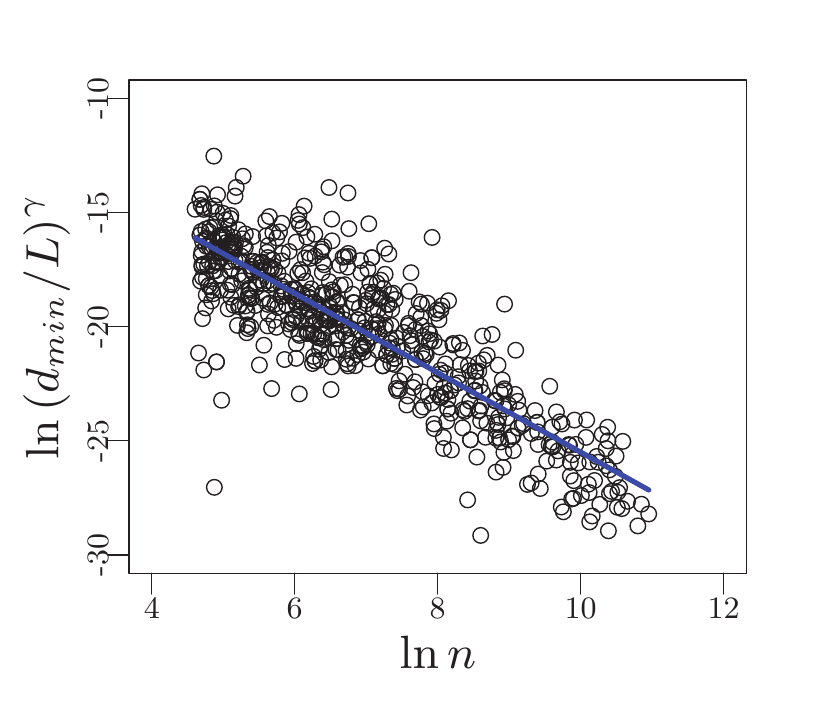}
  
  \end{subfigure}
\caption{In \ref{fig:dmingamvsn} the distribution of competitors $p(d)$ characterized by $d_{min}^{\gamma}$ is plotted against the number of buildings, while in \ref{fig:DistVsNbLnormalized} the normalized coexistence parameter is plotted against $n$ both on a log-log scale.   }
\end{figure}

The cumulative distributions of our cities $P(d)$ followed a power law whose lower cutoff we denote by $d_{min}$ and 
 exponent 
$\gamma+1$, which turns out to be clustered at 2.9 and 1.8. This correspond $p(d)$ with  $\gamma = $ 1.9 and 0.8 respectively as shown in Figure  \ref{gammavsnbbui}. 
The lower cluster is commensurate with the results of forest trees' scaling laws of \cite{Simini:2010jza}, which we suspect to be an ``organic setting" where species intermix in space whereas the higher value of $\gamma$ is a more ``discriminatory arrangement". To validate our claim we follow 
$\phi = d_{min}^{\gamma}$ as a function of $n$ as shown in Figure \ref{fig:dmingamvsn}. 

The red dashed line is the threshold $n_c$. Beyond that, the bifurcation curve's upper branch, with high values of $\phi$ compared to the lower branch,  corresponds to a setting where competing species are distant. Large $\phi$ can be traced back to, although not solely, a large value of $d_{min}$ which is indicative of local spatial homogeneity in building size where inter-competitor distance is large or is due to a large city extent equally leading to high $d_{min}$. The first is confirmed in the example of Arcadia and Tuscon, where the buildings sizes cluster homogeneously in space leading to a high value of $d_{min}$; they both belong to the upper branch of $\phi$. Conversely, Miami and Largo, belonging to the lower branch of $\phi$, exhibit high spatial mixing between species reflected by a low value of $d_{min}$. Their buildings' footprints are shown in the Supplementary Material in  {Figures S{(1-8)}}.
Additionally, the effect of city size is measured by renormalizing $d_{min}$ by $L$, where $L$ is the city's street length; it serves as a proxy to inter-buildings' distance. $(d_{min}/L)^{\gamma}$ exhibits a linear dependence on $n$ on a log-log scale as shown in Figure \ref{fig:DistVsNbLnormalized},
which confirms the additional dependence of the branchings on city size.

\begin{figure}
  \begin{subfigure}{ \label{fig:SizeEntropyVsNb}}
    \centering\includegraphics[width=7cm]{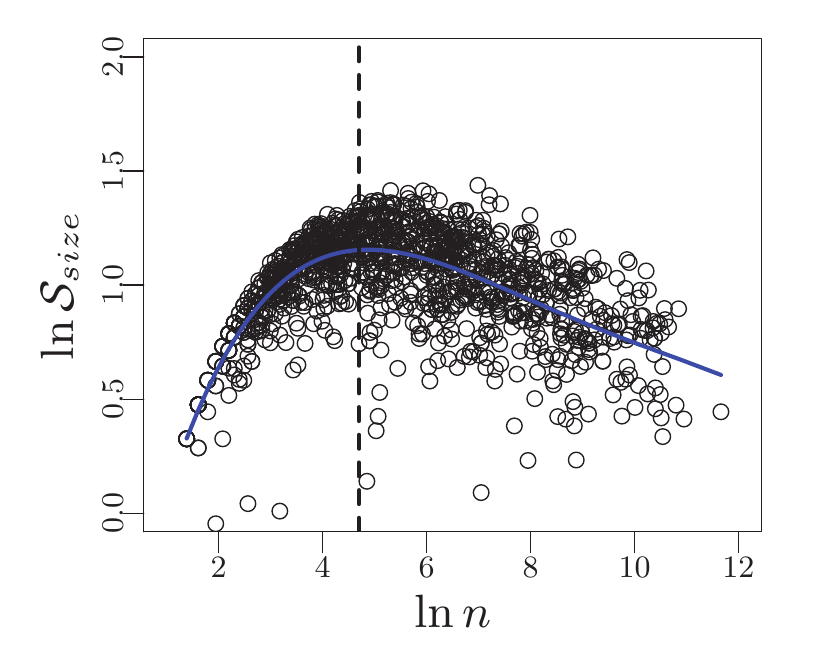}
   
  \end{subfigure}
  \begin{subfigure}{\label{fig:OrientationEntropyVsNb}}
    \centering\includegraphics[width=7cm]{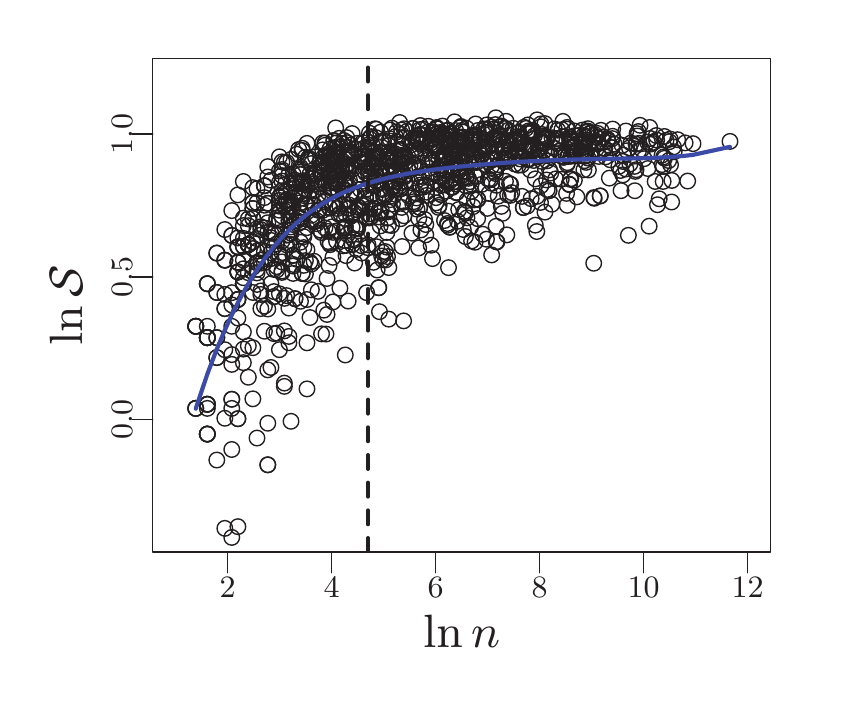}
    
  \end{subfigure}
  \caption{In (a) the buildings' size entropy versus $n$ is shown and in (b) the buildings' orientation entropy is plotted against $n$ both on a log-log scale. }
\end{figure}
\begin{figure}[!htp]%
 \center
    \includegraphics[width=0.38\textwidth]{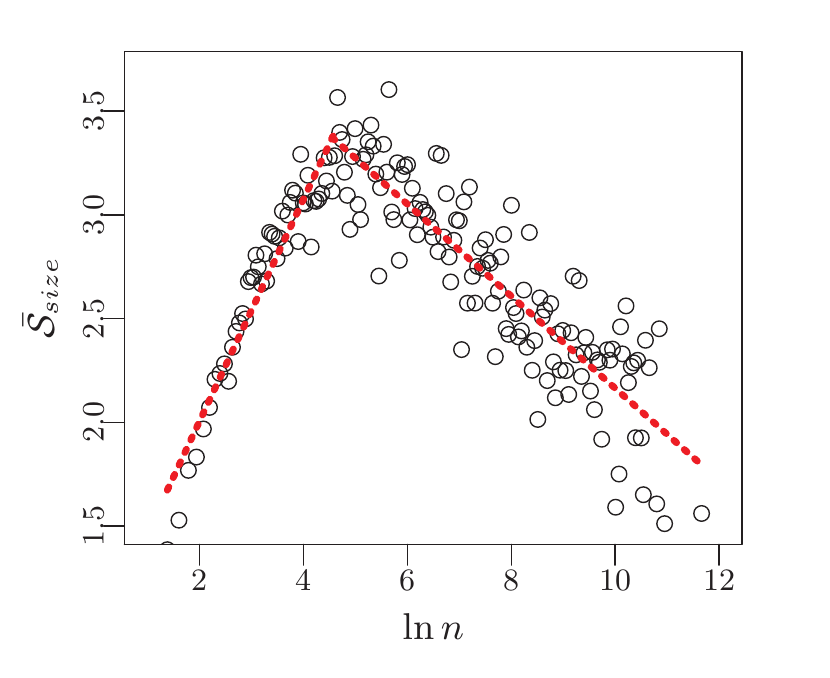} 

  \caption{ The linear models fitting the data of  $\bar{\mathcal{S}}_{size}$   versus $\log{n}$ are shown in red are respectively given by: $0.92+0.54\log{n}$ below $n_c$ and $4.38 -0.22\log{n}$ above it.    }
  {\label{fig:firstorder}}
\end{figure}

Further, the behavior of buildings' size and orientation entropies as a function of $n$, are shown in Figures \ref{fig:SizeEntropyVsNb} and \ref{fig:OrientationEntropyVsNb} 
respectively. 
It reveals that the buildings' size entropy $\mathcal{S}_{size}$, is maximized at $n_c$, beyond which the constraints on allowed buildings sizes increase; this reflects a tendency towards size homogeneity. For this range the buildings' orientation entropy  $\mathcal{S}$ is also near constant.  Below $n_c$, $\mathcal{S}_{size}$ and $\mathcal{S}$ increase, which corresponds to sparse cities with small number of buildings free to orient along any direction with no constraints on their sizes. Moreover, $\bar{\mathcal{S}}_{size}$ is plotted on a semi-logarithmic scale as shown in Figure \ref{fig:firstorder}.  

 We note that the ${d\mathcal{\bar{S}}_{size}}/{dn} = -{\mu}/{T}$, where $\mu$ is the chemical potential and $T$ is the city's ``temperature"; it is given by:

 \begin{equation}
{d\mathcal{\bar{S}}_{size}}/{dn} = -{\mu}/{T} = \begin{cases}
\ 0.54 \log{n} + 0.92 &n<n_c\\
-0.22\log{n} + 4.38 & n\ge n_c
\end{cases}
\end{equation}


 Since $\mu$ measures the necessary work to change the number of  ``particles", in this case buildings, by $dn$, or equivalently the system's resistance to adding an extra building, we conclude that below $n_c$, when $\frac{d\mathcal{\bar{S}}_{size}}{dn}$ increases, $ \frac{\mu}{T}$ decreases; beyond that the city becomes resistant to the addition of buildings; that is to say the construction of more buildings injects order into the city, which might be local or global.


This discontinuity in entropy is a signature of a second-order phase transition, differentiating between ``nascent-planning-free towns" and  ``planned cities" in both their mixed and segregated states, corresponding to both branches of $\phi$.  However,  the observed increase in orientation entropy beyond $n_c$ does not reflect the spatial distribution of order and thus is not able to reflect the difference between the upper and lower branches of $\phi$.
 
 \section*{ACKNOWLEDGEMENTS}
 
 M.G. is supported by the Natural Sciences and Engineering Research Council of Canada and by le Fonds de recherche du Qu\'ebec -- Nature et technologies (FRQNT).
 
\bibliography{scibib}
\end{document}